\documentclass[manuscript]{aastex}

\shorttitle{Reevaluation of Layered-Convection}
\shortauthors{Kurokawa and Inutsuka}

\begin{document}

\title{On the Radius Anomaly of Hot Jupiters: Reexamination of the Possibility and Impact of Layered Convection}

\author{Hiroyuki Kurokawa\altaffilmark{1,2}}
\email{hiro.kurokawa@elsi.jp}

\and

\author{Shu-ichiro Inutsuka\altaffilmark{2}}

\altaffiltext{1}{Earth-Life Science Institute, Tokyo Institute of Technology, 2-12-1-IE-10, Ookayama, Meguro-ku, Tokyo 152-8550, Japan}
\altaffiltext{2}{Department of Physics, Nagoya University, Furo-cho, Chikusa-ku, Nagoya, Aichi 464-8602, Japan}

\begin{abstract}
Observations have revealed that a significant number of hot Jupiters have anomalously large radii.
Layered convection induced by compositional inhomogeneity has been proposed to account for the radius anomaly of hot Jupiters.
To reexamine the impact of the compositional inhomogeneity, we perform an evolutionary calculation by determining convection regime at each evolutionary time step according to the criteria from linear analyses.
It is shown that the impact is limited in the case of the monotonic gradient of heavy element abundance.
The layered convection is absent for the first $1\ {\rm Gyr}$ from the formation of hot Jupiters and instead overturning convection develops.
The super-adiabaticity of the temperature gradient is limited by the neutrally stable state for the Ledoux stability criterion.
The effect of the increased mass of heavy elements essentially compensates the effect of the delayed contraction on the planetary radius caused by compositional inhomogeneity.
In addition, even in the case where the layered convection is artificially imposed, this mechanism requires extremely thin layers ($\sim 10^1-10^3\ {\rm cm}$) to account for the observed radius anomaly.
The long-term stability of such thin layers remains to be studied.
Therefore, if the criteria adopted in this paper are adequate, it might be difficult to explain the inflated radii of hot Jupiters by monotonic gradient of heavy element abundance alone.
\end{abstract}


\section{Introduction}

Masses and radii are fundamental quantities to constrain the bulk compositions of exoplanets as increasing the mass fraction of heavy elements in principle increases the density, 
and their compositions are naturally tied with their formation histories.
However, observations have revealed that a significant number of close-in gaseous planets (hot Jupiters) have anomalously large radii compared with the theoretical prediction for planets composed of hydrogen and helium \citep[e.g.,][]{Baraffe+2010,Baraffe+2014}.
Because the effect of increasing the mass fraction of heavy elements on density can veil that of the unknown mechanism of the radius anomaly, 
the mechanism might have influenced other exoplanets whose radius anomaly cannot be directly recognized.
Therefore, the radius anomaly may disables us from accurately determining not only the compositions of inflated hot Jupiters but also those of other exoplanets.

Several physical mechanisms have been proposed to account for the inflated radii of hot Jupiters.
The ideas can be classified into three categories \citep{Weiss+2013,Baraffe+2014}: incident stellar flux-driven mechanisms \citep[e.g.,][]{Showman+Guillot2002,Batygin+Stevenson2010,Arras+Socrates2010,Youdin+Mitchell2010}, tidal mechanisms \citep[e.g.,][]{Bodenheimer+2001,Gu+2003,Leconte+2010}, and delayed contraction \citep[e.g.,][]{Burrows+2007,Chabrier+Baraffe2007}.
Increasing statistics show a correlation between the incident stellar flux and the radius anomaly \citep{Demory+Seager2011, Weiss+2013}, 
yet none of the mechanisms has received a consensus \citep{Baraffe+2014}.

Delayed contraction due to layered convection in their interiors has been proposed to account for the radius anomaly \citep{Chabrier+Baraffe2007}.
Compositional inhomogeneity possibly inhibits large-scale-overturning convection and instead forms small-scale-layered convection, which is separated by diffusive interfaces \citep[e.g.,][]{Radko2003,Radko2005,Noguchi+Niino2010a,Noguchi+Niino2010b,Rosenblum+2011,Mirouh+2012,Wood+2013}.
Inefficient heat transport of the layered convection creates a super-adiabatic temperature profile, which causes the delayed contraction.
\citet{Chabrier+Baraffe2007} considered inhomogeneous internal profiles of heavy elements and demonstrated that the effect of the layered convection is sufficient to reproduce the radius anomaly by assuming the layered convection in the interiors of hot Jupiters.
In addition, \citet{Leconte+Chabrier2012,Leconte+Chabrier2013} proposed that the solar-system gas giants might be \textquotedblleft inflated" by this mechanism compared with the standard overturning-interior-models {that have the same masses of heavy elements.
Their layered-interior-model predicted the heavy-element enrichment of our gaseous giants that are up to $30\ \%$ to $60\ \%$ higher than previously thought \citep{Leconte+Chabrier2012} and successfully explained Saturn's luminosity problem \citep{Leconte+Chabrier2013}.
It is, therefore, crucial to evaluate the possibility and impact of the layered convection for the estimate of compositions of both the solar-system gas giants and extrasolar planets, and hence, crucial for constraining their origins.

While a simple structure of layered interior has been assumed to study its impacts \citep{Chabrier+Baraffe2007,Leconte+Chabrier2012,Leconte+Chabrier2013}, 
linear stability analyses and recent numerical simulations have shown that the layers form only in a limited parameter range described by the reciprocal of the density ratio $R_{\rho}^{-1}$, which is a function of both the temperature and mean-molecular-weight gradients \citep[][see Section \ref{sec_model} for details]{Rosenblum+2011,Mirouh+2012,Wood+2013,Leconte+Chabrier2012}.
The system is unstable for the large-scale-overturning convection when $R_{\rho}^{-1}$ is small, namely, when the destabilizing temperature gradient is too large.
A self-consistent treatment of the convection regimes is necessary to examine the possibility of the layered convection.

In addition, transport property of the layered convection depends significantly on the layer thickness.
Numerical simulations have found that layers successively merge and that merger of layers is accompanied by a significant increase in heat and compositional fluxes \citep{Radko2005,Noguchi+Niino2010b,Rosenblum+2011,Mirouh+2012,Wood+2013}.
It is still unknown whether there is an equilibrium layer thickness: \citet{Radko2005} discussed that the merger stops at an equilibrium thickness, whereas \citet{Noguchi+Niino2010b} stated that there seems to be no stable steady configuration of layers.
\citet{Chabrier+Baraffe2007} treated the layer thickness as an input parameter and showed that the impact of the layered convection on the radii of hot Jupiters depends on the layer thickness.
\citet{Leconte+Chabrier2012} demonstrated that the super-adiabaticity, and consequently, the estimate of heavy-element masses of Jupiter and Saturn depends on the layer thickness and \citet{Leconte+Chabrier2013} showed that the layer thickness affects the thermal evolution of Saturn.

The aim of our study is to reevaluate the possibility and impact of layered convection on the radii of hot Jupiters.
We will perform an evolutionary calculation of hot Jupiters with the self-consistent treatment of convection regimes.
The possibility of the layered convection due to the internal inhomogeneity of heavy-element abundance and the impact of the layer thickness will be studied.
Consequently, we will show that the possibility and impact of the layered convection are limited in the case of the monotonic gradient of heavy-element abundance and that it may be hard to account for the radius anomaly of hot Jupiters by the delayed contraction due to the layered convection alone.

\section{Model} \label{sec_model}

\subsection{Structure calculation}

We update the model of thermal evolution of exoplanets developed in \citet{Kurokawa+Kaltenegger2013} and \citet{Kurokawa+Nakamoto2014} for the description of non-adiabatic interior structures.
We calculate the thermal evolution of the interior structures of hot Jupiters with the Henyey method \citep[e.g.,][]{Kippenhahn+2012}.
The method solves the equations of the one-dimensional interior structure under hydrostatic equilibrium for the variables $P(M_r), T(M_r), r(M_r), L(M_r)$: 
\begin{equation}
\frac{\partial P}{\partial M_r} = -\frac{G M_r}{4 \pi r^4},
\end{equation}
\begin{equation}
\frac{\partial T}{\partial M_r} = -\frac{G M_r}{4 \pi r^4} \frac{T}{P} \nabla_T,
\end{equation}
\begin{equation}
\frac{\partial r}{\partial M_r} = \frac{1}{4 \pi r^2 \rho},
\end{equation}
\begin{equation}
\frac{\partial L}{\partial M_r} = -T\frac{{\rm d} S}{{\rm d} t}, 
\end{equation}
where $M_r$ is the enclosed mass, $P$ is the pressure, $T$ is the temperature, $r$ is the distance from the center of the planet, $L$ is the luminosity, $\rho$ is the density, $S$ is the entropy, $\nabla_T \equiv {\rm d} \ln{T}/ {\rm d} \ln{P}$, $t$ is the time, and $G$ is the gravitational constant.
The convection models give $\nabla_T$ as a function of the variables.

We use the analytical model of the irradiated atmosphere \citep{Guillot2010} and the Rosseland mean opacity tabulated by \citet{Freedman+2008}.
A power-law dependence fitted by \citet{Rogers+Seager2010} is used for the range out of the opacity table.
In the deep interior ($> 1\ {\rm Mbar}$), the conductive opacity calculated by \citet{Potekhin1999} is used.
The equation of state (EOS) of hydrogen and helium is taken from \citet{Saumon+1995} and the EOS of heavy elements (so-called \textquotedblleft metals") is represented by SESAME EOS for water \citep{Lyon+Johnson1992}.
Because the ionization degree is not provided in SESAME EOS, we assume the fully neutral state to maximize the stabilizing mean-molecular-weight gradient (i.e., to maximize the impact on the delayed contraction).
The mixing entropy is calculated with the expression summarized by \citet{Baraffe+2008}.
The approximate formula for the viscosity of hydrogen-helium mixture, $\nu = 4 \times 10^{-3}\ T_4^{-1/2}\ {\rm cm^2\ s^{-1}}$, is used, where $T_4$ is the temperature in units of $10^4\ {\rm K}$ \citep{Stevenson+Salpeter1977a}.
We use the diffusion coefficient of heavy-elements in hydrogen-helium mixture $D = 10^{-3}\ {\rm cm^2\ s^{-1}}$ as an order-of-magnitude estimate \citep{Stevenson+Salpeter1977a}.

\subsection{Convection regimes}

The energy is transported by radiation, conduction, and convection.
The convection regime is determined according to the criteria from linear stability analyses \citep{Ledoux1947,Walin1964,Kato1966}. 
The classification is based on the reciprocal of the density ratio,
\begin{equation}
R_{\rho}^{-1} \equiv \frac{\alpha_{\mu} \nabla_\mu}{\alpha_T (\nabla_{\rm d} - \nabla_{\rm ad})}, \label{eq_Rrho}
\end{equation}
where $\alpha_T \equiv -(\partial \ln{\rho} / \partial \ln{T})_{P,\mu}$, $\alpha_\mu \equiv (\partial \ln{\rho} / \partial \ln{\mu})_{P,T}$, $\nabla_{\rm ad} \equiv (\partial \ln{T}/ \partial \ln{P})_{S,\mu}$, $\nabla_{\mu} \equiv {\rm d} \ln{\mu}/ {\rm d} \ln{P}$, $\mu$ is the mean molecular-weight, and $\nabla_{\rm d}$ is {\it the temperature gradient needed to transport all energy by radiation or conduction} \citep{Rosenblum+2011,Mirouh+2012,Wood+2013,Leconte+Chabrier2012}.

\subsubsection{Stable regime}

When $R_{\rho}^{-1} < 0$ or $({\rm Pr}+1)/({\rm Pr}+\tau) < R_{\rho}^{-1}$, the energy is transported by radiation transfer or heat conduction, where ${\rm Pr} \equiv \nu/\kappa_T$ is the Prandtl number, $\tau \equiv D/ \kappa_T$ is the ratio of the compositional to heat diffusivities, and $\kappa_T$ is the heat diffusivity.
Under the assumption of the diffusion approximation for radiation transfer, both radiative and conductive temperature gradients are given by $\nabla_T = \nabla_{\rm d}$: 
\begin{equation}
\nabla_{\rm d} = \frac{3}{16 \pi a c G} \frac{\kappa L P}{M_r T^4},
\end{equation}
where $a$ is the radiation constant, $c$ is the speed of light in vacuum, and $\kappa$ is the radiative or conductive opacity.

\subsubsection{Layered-convection regime}

When $1 < R_{\rho}^{-1} < ({\rm Pr}+1)/({\rm Pr}+\tau)$, the diffusive instability leads to layered convection or turbulent diffusion \citep{Rosenblum+2011,Mirouh+2012,Wood+2013}.
Whereas recent numerical simulations showed that the layers form only in a limited parameter range within $1 < R_{\rho}^{-1} < ({\rm Pr}+1)/({\rm Pr}+\tau)$ \citep{Mirouh+2012}, 
we decided to assume the layered convection for $1 < R_{\rho}^{-1} < ({\rm Pr}+1)/({\rm Pr}+\tau)$ because the threshold $R_{\rho}^{-1}$ value is not given by an analytical formula.
We use a coarse graining model developed by \citet{Leconte+Chabrier2012} for the layered convection.
Instead of resolving fine-scale convective-layers and diffusive interfaces, the mean temperature gradient $\langle \nabla_T \rangle$ in the layered-convective layer is estimated.
The temperature gradient $\nabla_T$ is given by the mean temperature gradient $\langle \nabla_T \rangle$, namely, $\nabla_T = \langle \nabla_T \rangle$:
\begin{equation}
\langle \nabla_T \rangle \equiv \frac{\delta_T}{l+\delta_T} \nabla_{\rm d} + \frac{l}{l+\delta_T} \nabla_{T,l}, \label{eq_meanT}
\end{equation}
where $l$ is the layer thickness, $\delta_T$ is the interface thickness, and $\nabla_{T,l}$ is the temperature gradient within the layers.
In their model, $\nabla_{T,l}$ is calculated by assuming the relation, 
\begin{equation}
N\!u_T = Ra_{\rm \bigstar}^a, \label{eq_NuTRa}
\end{equation}
where $Ra_{\rm \bigstar}$ is a modified Rayleigh number and $N\!u_T$ is the thermal Nusselt number: 
\begin{equation}
Ra_{\rm \bigstar} \equiv \frac{\alpha_T g l^4}{\kappa_T^2 H_P} (\nabla_{T,l} - \nabla_{\rm ad}), \label{eq_Ra}
\end{equation}
\begin{equation}
N\!u_T \equiv \frac{\nabla_{\rm d} - \nabla_{\rm ad}}{\nabla_{T,l} - \nabla_{\rm ad}}, \label{eq_NuT}
\end{equation}
where $g$ is the gravitational acceleration and $H_P$ is the pressure scale height.
Equations \ref{eq_NuTRa}-\ref{eq_NuT} give,
\begin{equation}
\nabla_{T,l} = \nabla_{\rm ad} + (\nabla_{\rm d} - \nabla_{\rm ad}) \times Ra_{\rm \bigstar}^{-a}.
\end{equation}
Balancing the convective time scale in the layers and the thermal diffusive time scale in the interfaces leads to,
\begin{equation}
\frac{\delta_T}{l} = Ra_{\rm \bigstar}^{-\frac{1}{4}}. \label{eq_deltaT}
\end{equation}
Once the local variables $P,T,r,L$ are given, the mean temperature gradient in layered-convective zone $\langle \nabla_T \rangle$ is obtained from Equations (\ref{eq_meanT})-(\ref{eq_deltaT}).

The layered-convection model has two parameters: the exponent $a$ of the relation between $Ra_{\rm \bigstar}$ and $N\!u_T$ (Equation (\ref{eq_NuTRa})) and the ratio of the layer thickness to the pressure scale height $l/H_P$.
In our model, we assume $a=0.5$, the value consistent with the mixing length theory.
The layer thickness in quasi-steady state is unknown \citep[e.g.,][]{Rosenblum+2011}.
The lower limit is given by the thickness of the layers initially formed after the saturation of the diffusive instability.
Numerical simulations \citep[e.g.,][]{Rosenblum+2011} showed that the initial layer thickness is $\sim 10^{1}-10^{2}\ d$, where $d$ is the thermal diffusion scale \citep[e.g.,][]{Baines+Gill1969}, which is given by,
\begin{equation}
d = \Biggl( \frac{\kappa_T \nu}{\alpha_T g^2 \rho/P (\nabla_T - \nabla_{\rm ad})} \Biggr)^{\frac{1}{4}}. \label{eq_d}
\end{equation}
Assuming typical values, $\kappa_T \sim 10^{-1}-10^{0}\ {\rm cm^2}\ {\rm s^{-1}}$, $\nu \sim 10^{-3}-10^{-2}\ {\rm cm^2}\ {\rm s^{-1}}$, $\alpha_T \sim 1$, $g \sim 10^3\ {\rm cm}\ {\rm s^{-2}}$, $\rho \sim 1\ {\rm g}\ {\rm cm^{-3}}$, and $P \sim 1\ {\rm Mbar}$ \citep{Chabrier+Baraffe2007,Leconte+Chabrier2012}, Equation (\ref{eq_d}) leads to $d^4 \sim 10^{2-4}\ (\nabla_T - \nabla_{\rm ad})$.
The temperature gradient in the layers-convective zone can be estimated from the Ledoux criterion as $\nabla_T - \nabla_{\rm ad} \sim \alpha_\mu/\alpha_T \nabla_\mu$.
Assuming $\alpha_\mu \sim 1$ and $\nabla_\mu \sim 1$, $d$ is estimated to be $\sim 10^{0.5}-10^{1}\ {\rm cm}$.
Therefore, the initial layer thickness is estimated to be $l \sim 10^{1}-10^{2}\ d \sim 10^{1.5}-10^{3}\ {\rm cm}$.
We use $l/H_P \sim l/R_p \sim 10^3\ {\rm cm}/10^{10}\ {\rm cm} \sim 10^{-7}$ in our nominal model, where $R_p$ is the planetary radius.

\subsubsection{Overturning-convection regime}

The system is unstable for the overturning convection when $0 < R_{\rho}^{-1} < 1$.
We derive a heat transport model for the overturning convection in the presence of the mean-molecular-weight gradient by extending the standard mixing-length theory \citep[e.g.,][]{Kippenhahn+2012}.
We suppose that a convective element conserves its composition until the mixing with surroundings.
The total energy flux $L/4\pi r^2$ at a given point consists of the conductive flux $F_{\rm d}$ and the convective flux $F_{\rm conv}$.
The sum can be written as,
\begin{equation}
F_{\rm d} + F_{\rm conv} = \frac{4acG}{3} \frac{T^4 M_r}{\kappa P r^2} \nabla_{\rm d}. \label{eq_F}
\end{equation}
The conductive flux $F_{\rm d}$ is given by,
\begin{equation}
F_{\rm d} = \frac{4acG}{3} \frac{T^4 M_r}{\kappa P r^2} \nabla_T.\label{eq_Fd}
\end{equation}
The convective flux $F_{\rm conv}$ is given by, 
\begin{equation}
F_{\rm conv} = \rho v_{\rm e} C_P T (\nabla_T - \nabla_{T,{\rm e}})\frac{1}{2}\frac{l_{\rm m}}{H_P},\label{eq_Fconv}
\end{equation}
where $v_{\rm e}$ and $\nabla_{T,{\rm e}}$ are the velocity and the temperature gradient of the convective element, $C_P$ is the heat capacity, and $l_{\rm m}$ is the mixing length, respectively.
The velocity is estimated from the work done by buoyancy force as,
\begin{equation}
v_{\rm e}^2 = g (\alpha_T \nabla_T - \alpha_{\mu} \nabla_{\mu} - \alpha_T \nabla_{T,{\rm e}}) \frac{1}{8} \frac{l_{\rm m}^2}{H_P}.\label{eq_ve}
\end{equation}
Using Equations (\ref{eq_F})-(\ref{eq_ve}), we obtain,
\begin{equation}
(\nabla_T - \nabla_{T,{\rm e}}) \Biggl( \nabla_T - \nabla_{T,{\rm e}} - \frac{\alpha_{\mu}}{\alpha_T} \nabla_{\mu} \Biggr)^{\frac{1}{2}} = \frac{8}{9} U(\nabla_{\rm d} - \nabla_{T}), \label{eq_MLT1}
\end{equation}
where,
\begin{equation}
U \equiv \frac{3acT^3}{C_P \rho^2 \kappa l_{\rm m}^2} \Biggl( \frac{8 H_P}{g \alpha_T} \Biggr)^{\frac{1}{2}}.
\end{equation}
Considering the conductive energy loss of the element gives \citep{Kippenhahn+2012}, 
\begin{equation}
\nabla_{T,{\rm e}} - \nabla_{\rm ad} = \frac{6acT^3}{\kappa \rho^2 C_P l_m v_{\rm e}}. \label{eq_Te}
\end{equation}
We substitute Equation (\ref{eq_ve}) for Equation (\ref{eq_Te}) and obtain,
\begin{equation}
(\nabla_{T,{\rm e}} - \nabla_{\rm ad}) \Biggl( \nabla_T - \nabla_{T,{\rm e}} - \frac{\alpha_{\mu}}{\alpha_T} \nabla_{\mu} \Biggr)^{\frac{1}{2}} = 2U (\nabla_T - \nabla_{T,{\rm e}}). \label{eq_MLT2}
\end{equation}
Finally, Equations (\ref{eq_MLT1}) and (\ref{eq_MLT2}) lead to,
\begin{equation}
X(X-W) + \frac{1}{2U} X (X-W)^{\frac{3}{2}} -\frac{Y}{2U} (X-W)^{\frac{3}{2}} + \frac{9}{16U^2} X (X-W)^2 = 0,\label{eq_MLT-m}
\end{equation}
where $X \equiv \nabla_T - \nabla_{T,{\rm e}}$, $Y \equiv \nabla_{\rm d} - \nabla_{\rm ad}$, and $W \equiv (\alpha_{\mu} / \alpha_T) \nabla_{\mu}$, respectively.
Equation (\ref{eq_MLT-m}) is numerically solved for $X$.
Then the temperature gradient $\nabla_T$ is obtained by substituting $X$ for Equation (\ref{eq_MLT1}).
The model derived here agrees with the model derived in \citet{Stevenson+Salpeter1977b} when we approximate $\nabla_{T,{\rm e}}$ as $\nabla_{T,{\rm e}} = \nabla_{\rm ad}$.
A general form of the extended mixing-length theory was described by \citet{Umezu+Nakakita1988}.

\subsection{Settings}

We assume a Jupiter mass planet and the equilibrium temperature of $1250\ {\rm K}$.
The mean entropy of $10\ k_{\rm B}\ {\rm baryon^{-1}}$ is assigned for the initial state of the self-consistent convection models \citep{Marley+2007}. 
A lower value, $8.8\ k_{\rm B}\ {\rm baryon^{-1}}$, is assigned for the layered-convection models to avoid unrealistically large initial radii.
The initial temperature profile is calculated to satisfy the condition that the internal luminosity $L(M_r)$ linearly decreases from the intrinsic luminosity at the top to zero  at the center of the planet (i.e., the cooling rate is constant through the planet).

We calculate the evolution for three different compositional profiles: the metal-poor model, metal-rich model, and monotonic-gradient model (Figure \ref{fig_M-Z}).
The metal-poor model has protosolar elemental-abundance ($Y=0.28$, $Z=0.02$, where $Y$ and $Z$ are the abundances of helium and heavy elements) throughout the interior. 
The monotonic-gradient model has a gradient of heavy-element abundance within the inner $30\ \%$ by mass the same with the model of \citet{Chabrier+Baraffe2007}.
We use the monotonic-gradient model in Figure \ref{fig_M-Z} to study the impact of the compositional inhomogeneity in this paper except for Figure \ref{fig_Age-R_consistent_ad}.
The metal-rich model has the same mass of heavy elements with the monotonic-gradient model but the homogeneous distribution is assumed.
Because our model is aimed at determining convection regimes and calculating the evolution for given compositional profiles, the compositional evolution is not calculated for simplicity.
The evolution of the compositional profile will be discussed in Section \ref{sec_discussion}.

We calculate the evolution of hot Jupiters by using both the self-consistent convection model and the layered-convection model for comparison. 
The layered-convection model with the monotonic compositional gradient effectively corresponds to the case of staircase-like compositional profile studied by \citet{Chabrier+Baraffe2007}.

\section{Results} \label{sec_results}

\subsection{Possibility of the layered convection}

First, we reexamine the possibility of the layered convection in a similar setup with \citet{Chabrier+Baraffe2007}.
The assumed heavy-element profiles and the evolution of the radii of planets are shown in Figures \ref{fig_M-Z} and \ref{fig_Age-R_consistent}.
The planets initially have large radii and contract with time because of cooling.
The metal-rich model has a smaller radius than the metal-poor model because of the larger weight of heavy elements.
The monotonic-gradient model has an anomalously large radius compared with the homogeneous (metal-poor and metal-rich) models in the case where the layered convection is artificially assumed.
This model corresponds to the case of staircase-like compositional profile studied by \citet{Chabrier+Baraffe2007}.
The inefficient heat transport of the layered convection leads to the super-adiabatic temperature gradient in the interior (Figure \ref{fig_M-T_consistent}).
The delayed contraction causes the radius anomaly that matches the observation as shown by \citet{Chabrier+Baraffe2007}.

However, the impact of the compositional inhomogeneity is limited in the case where the self-consistent treatment of convection regimes is adopted (Figure \ref{fig_Age-R_consistent}).
Though the monotonic-gradient model has a slightly larger radius than the homogeneous metal-rich model in which the same mass of heavy elements is assumed, 
the radius is at most comparable with that of the homogeneous metal-poor model.
This means that the effect of the increased mass of heavy elements compensates the effect of the compositional inhomogeneity on the planetary radius.
As a result, the compositional inhomogeneity cannot reproduce the observed large radius anomaly (up to $\simeq 2$ Jupiter radius) in the setting of the present paper even if the compositional gradient was conserved.
The evolution of planetary radii calculated for different heavy-element profiles are shown in Figure \ref{fig_Age-R_consistent_ad}, where the self-consistent treatment of convection regimes is adopted.
The impact of the compositional inhomogeneity is insufficient to reproduce the inflated radii of hot Jupiters in all the cases.

The reason for the limited effect is the absence of the layered convection.
The reciprocal of the density ratio $R_{\rho}^{-1}$ in the interior is shown in Figure \ref{fig_M-Rrho_consistent}.
As mentioned in Section \ref{sec_model}, the convection regime is determined by the density ratio.
The convection regime is the overturning convection for the first $1\ {\rm Gyr}$.
In the overturning-convection regime, the efficient heat transport forces the temperature gradient to follow the neutrally stable state for the Ledoux criterion.
Consequently, the super-adiabaticity is limited as $\nabla_T \simeq \nabla_{\rm ad} + \alpha_\mu / \alpha_T \nabla_\mu$ (Figure \ref{fig_M-T_consistent}) and the effect on the delayed contraction is limited (Figure \ref{fig_Age-R_consistent}).

On the contrary, the layered-convection model leads to a higher internal temperature (Figure \ref{fig_M-T_consistent}).
The homogeneous models have adiabatic temperature profiles and their central temperature is $2-3\ \times 10^4\ {\rm K}$.
The layered-convection model has a super-adiabatic temperature profile in the inhomogeneous composition zone (the inner $30\%$ by mass).
The central temperature reaches $\simeq 2\ \times 10^5\ {\rm K}$ as shown by \citet{Chabrier+Baraffe2007}.
However, the internal temperature is lower in the case of the self-consistent convection model.
The layer forms after $1\ {\rm Gyr}$ passed, when the planet has already cooled (Figure \ref{fig_M-Rrho_consistent}).
The temperature gradient in the layered-convection regime never exceeds $\nabla_T = \nabla_{\rm ad} + \alpha_\mu / \alpha_T \nabla_\mu$ (Figure \ref{fig_M-T_consistent}).
The radius of the self-consistent convection model matches that of layered-convection model when the layered-convection zone sufficiently develops after $10\ {\rm Gyr}$ (Fig. \ref{fig_Age-R_consistent}).

\subsection{Dependence on the layer thickness}

Second, we study the dependence on the layer thickness. 
Because the thickness of the layered convection in quasi-steady state is poorly constrained \citep[e.g.,][]{Rosenblum+2011}, it is treated an input parameter in our model. 
Figures \ref{fig_Age-R_layered} and \ref{fig_M-T_layered} shows the evolution of the radii and the internal temperature profiles for the layered-convection models with the different layer thickness.
Hot Jupiters keep larger radii for smaller layer thickness (Figure \ref{fig_Age-R_layered}) as shown by \citet{Chabrier+Baraffe2007}.
This is because thinner layers result in inefficient heat transport and consequently lead to higher internal temperature (Figure \ref{fig_M-T_layered}).
There is an asymptotic upper limit of the radius (shown by the results for $l/H_P = 10^{-9}$ and $l/H_P = 10^{-8}$) caused by the upper limit of the mean temperature gradient in the layered-convective zone: $\nabla_T = \nabla_{\rm d}$.

Even in the case where the layered convection is assumed, the layer thickness of $l/H_P \sim 10^{-9} - 10^{-7}$ is necessary to account for the observed large anomaly.
As the scale height in the interior is comparable with the planetary radius ($H_P \sim R_p \sim 10^{10}\ {\rm cm}$), $l/H_P \sim 10^{-9} - 10^{-7}$ corresponds to $l \sim 10^{1}-10^{3}\ {\rm cm}$.
This value is comparable with the layer thickness that initially formed after the saturation of the diffusive instability ($l \sim 10^{1.5}-10^{3}\ {\rm cm}$, see Section \ref{sec_model}).
It is still unknown whether the layers can keep this small value after the layer mergers \citep[e.g.,][]{Rosenblum+2011}.

Figure \ref{fig_Age-R_consistent_l} shows the evolution of the radii and the internal temperature profiles for the self-consistent convection models with the different layer thickness.
The evolution before $1\ {\rm Gyr}$ does not depend on the layer thickness because the layered convection is absent (Figure \ref{fig_M-Rrho_consistent}).
The radii are almost independent of the layer thickness even after the layer formation ( $> 1\ {\rm Gyr}$), that is different from the results of the layered-convection models (Figure \ref{fig_Age-R_layered}).
Thinner layers result in inefficient heat transport and consequently lead higher internal temperature.
However, the effect of the temperature on the radii is weak because planets are sustained by the degeneracy pressure of electrons in this late stage. 
Therefore, the evolution of the radius poorly depends on the layer thickness in the case where the self-consistent treatment of the convection regimes is adopted.

\section{Discussion} \label{sec_discussion}

\subsection{Evolution of the compositional profile}

Our results showed that the overturning convection develops in the interiors of hot Jupiters for the first $1\ {\rm Gyr}$ and that layers form only in the late stage in the case of the monotonic compositional gradient (Section \ref{sec_model}).
The evolution of the compositional profile is not considered in our model, but the overturning convection may smooth out the compositional inhomogeneity efficiently.
Here, we estimate the effective diffusion coefficient $D_{\rm eff}$ of each convection regime and the mixing time scale $t_{\rm mix}$.
For the stable system, $D_{\rm eff}$ is purely $D_{\rm eff} = D$.
For the layered-convection, we estimate $D_{\rm eff}$ by considering the diffusion in the compositional interfaces, as,
\begin{equation}
D_{\rm eff} \sim \frac{l+\delta_{\rm z}}{\delta_{\rm z}} D,\label{eq_Deff1}
\end{equation}
where $\delta_{\rm z}$ is the thickness of the compositional interface, which can be estimated as $\delta_{\rm z} \sim \delta_T \tau^{1/2}$ \citep{Leconte+Chabrier2012}.
Equation (\ref{eq_Deff1}) can be written as,
\begin{equation}
D_{\rm eff} \sim \frac{Ra_{\rm \bigstar}^{\frac{1}{4}}+\tau^{\frac{1}{2}}}{\tau^{\frac{1}{2}}} D. \label{eq_Deff2}
\end{equation}
For the overturning convection, $D_{\rm eff}$ is estimated as,
\begin{equation}
D_{\rm eff} \sim \frac{1}{2} v_{\rm e} l_{\rm m}.
\end{equation}
The mixing time scale $t_{\rm mix}$ is estimated as $t_{\rm mix} \sim R_p^2 / D_{\rm eff}$ by using $D_{\rm eff}$.

The effective diffusion coefficient $D_{\rm eff}$ and the mixing time scale for composition $t_{\rm mix}$ are shown in Figure \ref{fig_M-D}.
There are three branches in the mixing time scale.
The shortest mixing time scale ($\sim 10^3\ {\rm yr}$) corresponds to the overturning-convection regime.
The branches of longer time scales that appear at $t >1\ {\rm Gyr}$ are those of the layered-convection regime and the stable regime.
The short mixing time scale suggests that the overturning convection in the early stage may smooth out any compositional inhomogeneity.
In contrast, the compositional profile is expected to be preserved in the layered-convection regime in the late stage.

\subsection{Relation to \citet{Chabrier+Baraffe2007}}

\citet{Chabrier+Baraffe2007} assumed layered-convective zones in the interiors of hot Jupiters and calculated their thermal evolution by resolving the layered-convective layers directly.
They concluded that the effect of the layered convection is sufficient to reproduce the inflated radii of hot Jupiters.
On the other hand, we determined convection regime according to the criteria from linear analyses by using the coarse graining model of the layered convection developed by \citet{Leconte+Chabrier2012} in which the layered structure is not resolved.
We showed that the effect is insufficient to explain the radius anomaly because of the formation of the overturning convection.
The difference of these two approaches is contrasted here.

Suppose a staircase-like, layered-convective zone where the coarse grained \textquotedblleft macroscopic" structure is unstable for the overturning convection (namely, unstable for the Ledoux stability criterion) and the \textquotedblleft local" structure in the diffusive interfaces is stable.
The model of \citet{Chabrier+Baraffe2007} adopted layered convection model for this configuration by judging from the \textquotedblleft local" stability.
In contrast, we adopted the overturning convection by judging from the \textquotedblleft macroscopic" stability. 
Fluid dynamical simulations are required to determine the long-term stability of the layered convection in the system where \textquotedblleft macroscopic" structure is unstable for the overturning convection.
If the layered convection state is unstable on a relatively short timescale ($< 1\ {\rm Gyr}$), the structure may evolve into overturning convection state.
In this case, our approach is more realistic.

\subsection{Other possibilities}

We discussed that the overturning convection may smooth out the compositional inhomogeneity based on the mixing timescale.
However, compositional transport of the overturning convection may possibly create a sharp, stabilizing compositional gradient before it is smoothed out.
\citet{Vazan+2015} found the formation of staircase-like compositional-profiles caused by the compositional transport.
This sharp compositional-gradient may preserve the inhomogeneity for billions of years.
Solving both the thermal and compositional evolution with the self-consistent treatment of the convection regimes are necessary to study this possibility.
It would eventually test the possibility to form a staircase-like compositional profile assumed by \citet{Chabrier+Baraffe2007} from a monotonic compositional gradient as well.
If fine-scale layers ($\sim 10^1-10^3\ {\rm cm}$) are formed in this stage, it may result in the delayed contraction enough to explain the radius anomaly.

Although compositional inhomogeneity created in the formation stage may be smoothed out by the overturning convention in the early stage,
compositional inhomogeneity that emerges in the late phase may contribute to form the layered convection in the interiors of giant planets.
Erosion of the core \citep{Guillot+2004,Wilson+Militzer2012a,Wilson+Militzer2012b} and phase separation of hydrogen and helium \citep{Salpeter1973,Stevenson1975,Stevenson+Salpeter1977a,Stevenson+Salpeter1977b,Nettelmann+2015}  are the possible mechanisms.
The acquired layered-convection may account for luminosity problems of solar system giant planets \citep{Leconte+Chabrier2013}, but it might be hard to account for the inflated radii of hot Jupiters by this acquired layered-convection alone.

Our results suggest that it is hard to explain the inflated radii of hot Jupiters by the compositional inhomogeneity alone at least in the case of the monotonic compositional-profiles.
As discussed by \citet{Baraffe+2010,Baraffe+2014}, the solution could be a combination of various processes.
If there is another mechanism to delay the contraction, 
it would help the formation of the layered convection by increasing the value of $R_{\rho}^{-1}$.
Additional energy source deposited in enough deep region \citep{Ginzburg+Sari2015} and atmospheric enhanced opacities \citep{Burrows+2007} are the possible mechanisms of delayed contraction.
It should be interesting to study the combination of the layered convection with other processes to account for the radius anomaly of hot Jupiters.

\section{Summary}

Layered convection induced by compositional inhomogeneity has been proposed to account for the inflated radii of hot Jupiters.
We developed an evolutionary model with a self-consistent treatment of convection regimes and applied the model to the hot Jupiters that have the monotonic compositional gradients.
The layered convection was absent for the first $1\ {\rm Gyr}$ and instead overturning convection developed in the interior.
Whereas the layered-convection model led to a higher internal temperature, the self-consistent convection model led to a relatively lower internal temperature.
As a result, the impact of the compositional inhomogeneity on the radius was limited.
Because the layered convection is absent in the early stage, the assumption on the layer thickness does not affect the evolution.
We concluded that it is hard to explain the inflated radii of hot Jupiters by the compositional inhomogeneity at least in the case of the monotonic compositional gradient.
Efficient mixing due to the overturning convection may smooth out the compositional inhomogeneity initially presented.
Further studies are needed to understand the consequences of the compositional transport.
Core erosion or phase separation may contribute the late formation of the compositional gradient and the layered convection.
The acquired layered-convection may account for luminosity problems of solar system giant planets, but it might be difficult to account for the inflated radii of hot Jupiters by this mechanism alone.

\section*{Acknowledgments}

HK thanks Takeru Suzuki for inputs to develop the Heyney code and Takashi Noguchi for fruitful discussion about double-diffusive convection.
This work was supported by Grants-in-Aid from the Japanese Ministry of Education, Culture, Sports, Science and Technology (MEXT) (23244027 and 23103005).
HK was supported by JSPS KAKENHI Grant Number 15J09448.

\clearpage

\begin{figure}
\begin{center}
\includegraphics[scale=1.0]{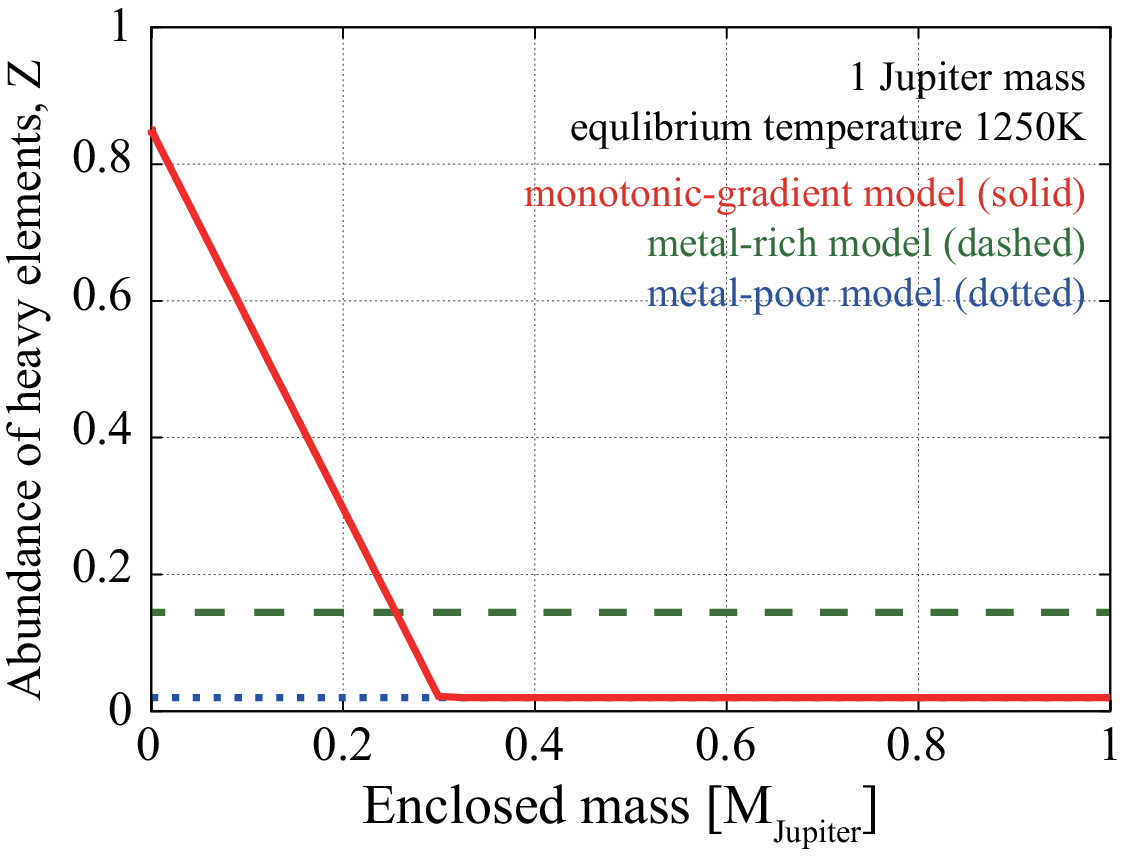}
\caption{
Assumed heavy-element profiles in the interiors of hot Jupiters for monotonic-gradient model (red solid line), homogeneous metal-rich model (green dashed line), and homogeneous metal-poor model (blue dotted line), respectively.
The monotonic-gradient model and the homogeneous metal-rich model have the same total mass of the heavy elements. 
\label{fig_M-Z}}
\end{center}
\end{figure}

\clearpage

\begin{figure}
\begin{center}
\includegraphics[scale=1.0]{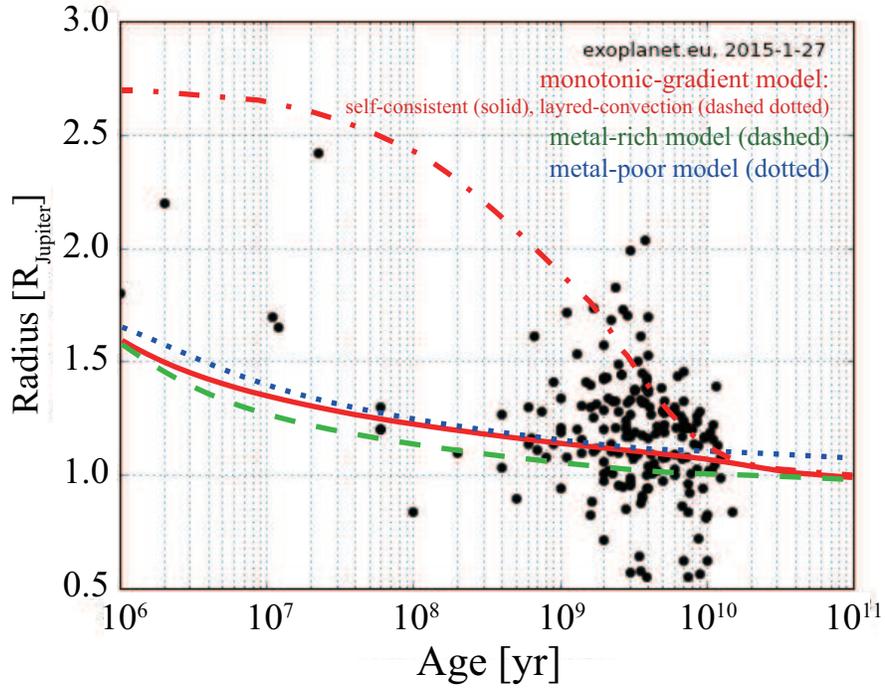}
\caption{
Evolution of the radii of hot Jupiters.
The results for the monotonic-gradient model calculated with self-consistent treatment of convection regimes (red solid line), the monotonic-gradient model calculated with the layered convection (red dashed-dotted line), the metal-rich model (green dashed line), and the metal-poor model (blue dotted line), are shown, respectively.
The assumed heavy-element profiles are shown in Figure \ref{fig_M-Z}. 
Filled circles indicate observed exoplanets taken from exoplanet.eu.
\label{fig_Age-R_consistent}}
\end{center}
\end{figure}

\clearpage

\begin{figure}
\begin{center}
\includegraphics[scale=1.0]{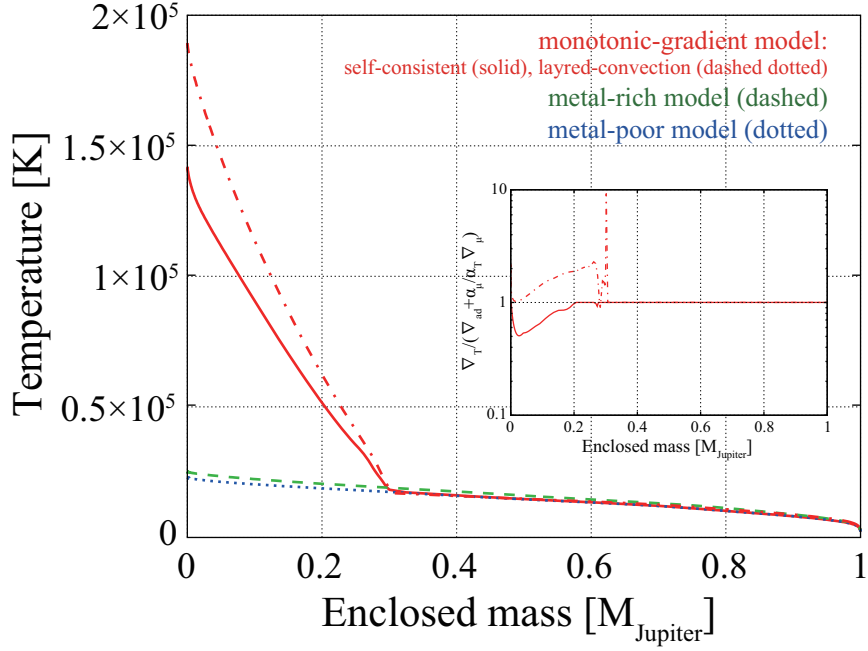}
\caption{
Internal temperature profiles at $5\ {\rm Gyr}$.
Colors and line types indicate the assumed heavy-element profiles and the convection models the same as Fig. \ref{fig_Age-R_consistent}.
Scaled temperature gradient profiles, $\nabla_T /( \nabla_{\rm ad} + \alpha_\mu / \alpha_T \nabla_\mu)$, are portrayed in the inner subset.
The results for the self-consistent convection model (red solid line) and the layered-convection model (red dashed-dotted line) are shown.
\label{fig_M-T_consistent}}
\end{center}
\end{figure}

\clearpage

\begin{figure}
\begin{center}
\includegraphics[scale=1.0]{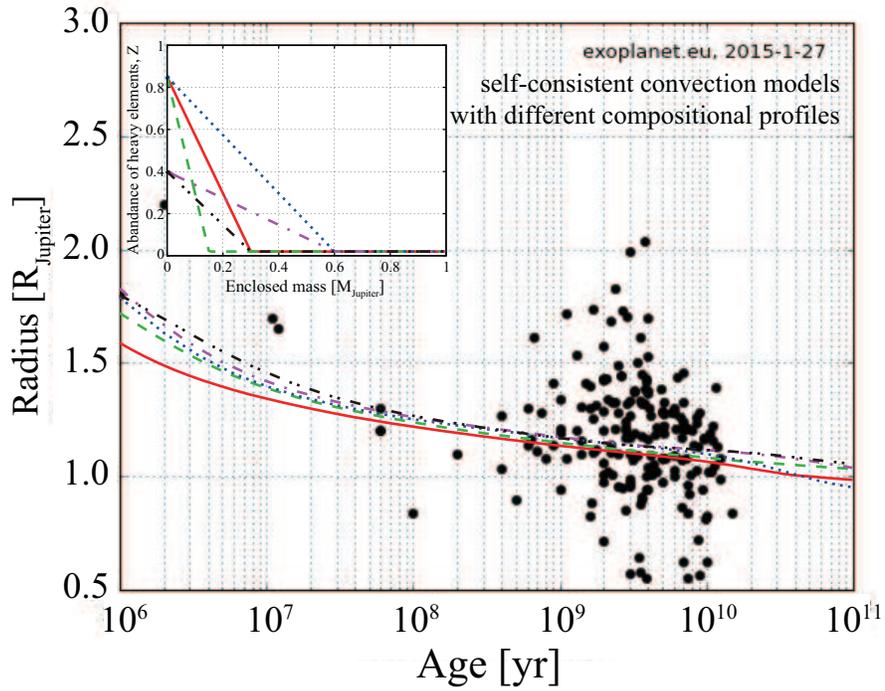}
\caption{
Evolution of the radii of hot Jupiters calculated for different heavy-element profiles. 
The self-consistent treatment of convection regimes is adopted.
The heavy-element profiles in the interiors are portrayed in the inner subset.
\label{fig_Age-R_consistent_ad}}
\end{center}
\end{figure}

\clearpage

\begin{figure}
\begin{center}
\includegraphics[scale=1.0]{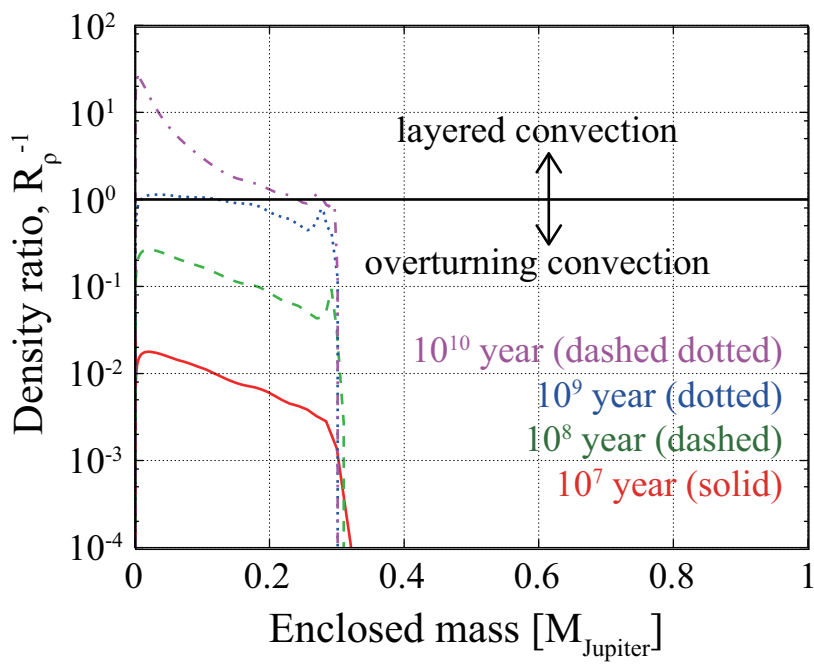}
\caption{
Density ratio profiles in the interior of the monotonic-gradient model calculated with the self-consistent treatment of convection regimes (the case of red solid line in Fig. \ref{fig_Age-R_consistent}).
The results are shown for $0.01\ {\rm Gyr}$ (red solid line), $0.1\ {\rm Gyr}$ (green dashed line), $1\ {\rm Gyr}$ (blue dotted line), and $10\ {\rm Gyr}$ (magenta dashed-dotted line), respectively.
\label{fig_M-Rrho_consistent}}
\end{center}
\end{figure}

\clearpage

\begin{figure}
\begin{center}
\includegraphics[scale=1.0]{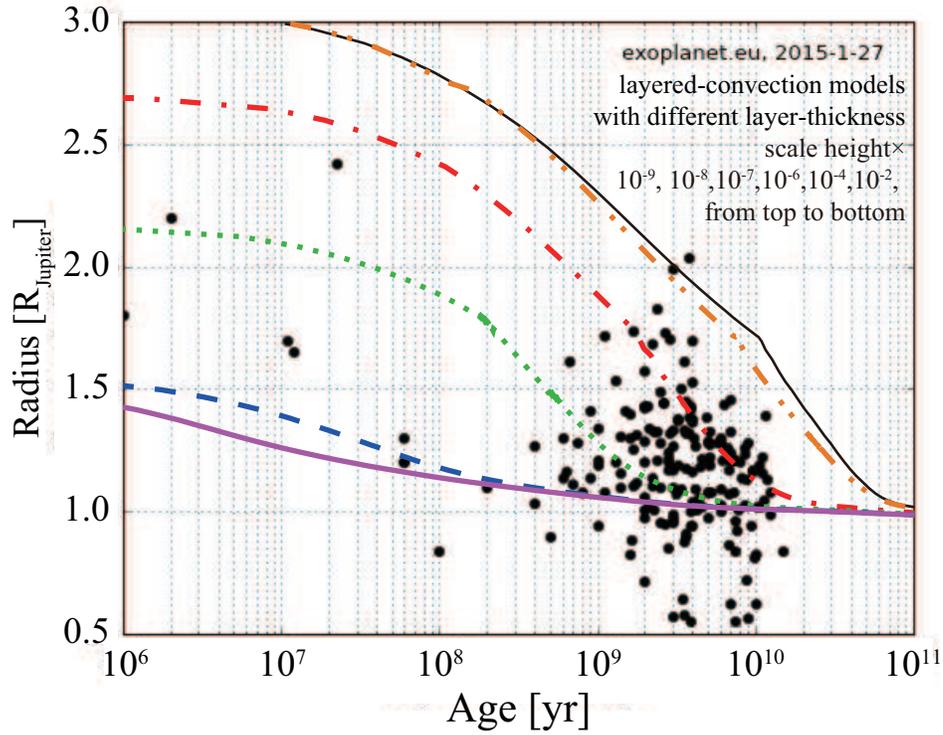}
\caption{
Evolution of the radii of hot Jupiters calculated by using the layered-convection model with the different layer thickness. 
The monotonic-gradient model in Figure \ref{fig_M-Z} is adopted for the profile of heavy-element abundance. 
The results are shown for $l/H_P = 10^{-9}$, $10^{-8}$, $10^{-7}$, $10^{-6}$, $10^{-4}$, and $10^{-2}$ from top to bottom, respectively.
\label{fig_Age-R_layered}}
\end{center}
\end{figure}

\clearpage

\begin{figure}
\begin{center}
\includegraphics[scale=1.0]{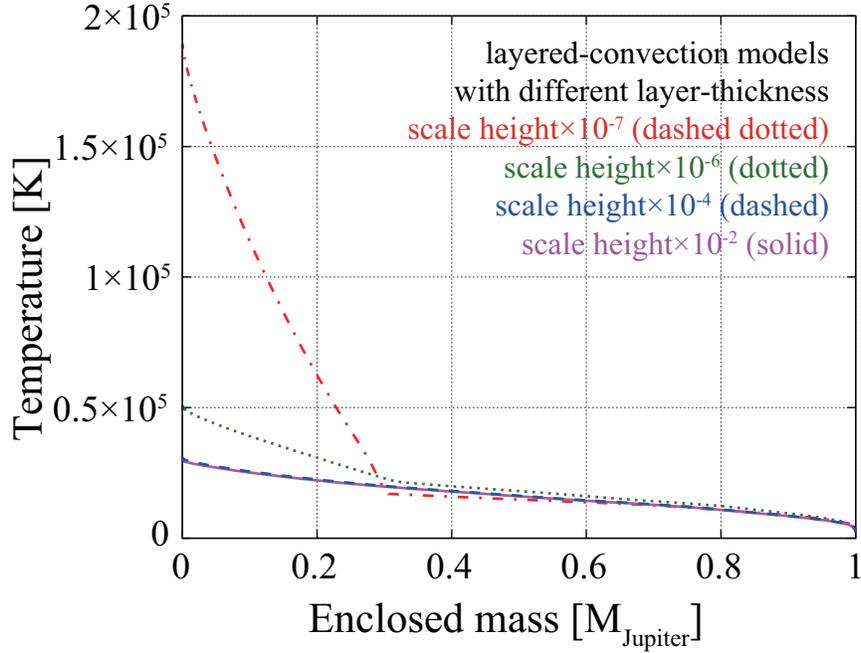}
\caption{
Internal temperature profiles at $5\ {\rm Gyr}$ of monotonic-gradient models where the layered convection is artificially assumed.
The monotonic-gradient model in Figure \ref{fig_M-Z} is adopted for the profile of heavy-element abundance.
Colors indicate the assumed layer thickness:  $l/H_P = 10^{-7}$ (red dashed-dotted line), $10^{-6}$ (green dotted line), $10^{-4}$ (blue dashed line), and $10^{-2}$ (magenta solid line), respectively.
\label{fig_M-T_layered}}
\end{center}
\end{figure}

\clearpage

\begin{figure}
\begin{center}
\includegraphics[scale=1.0]{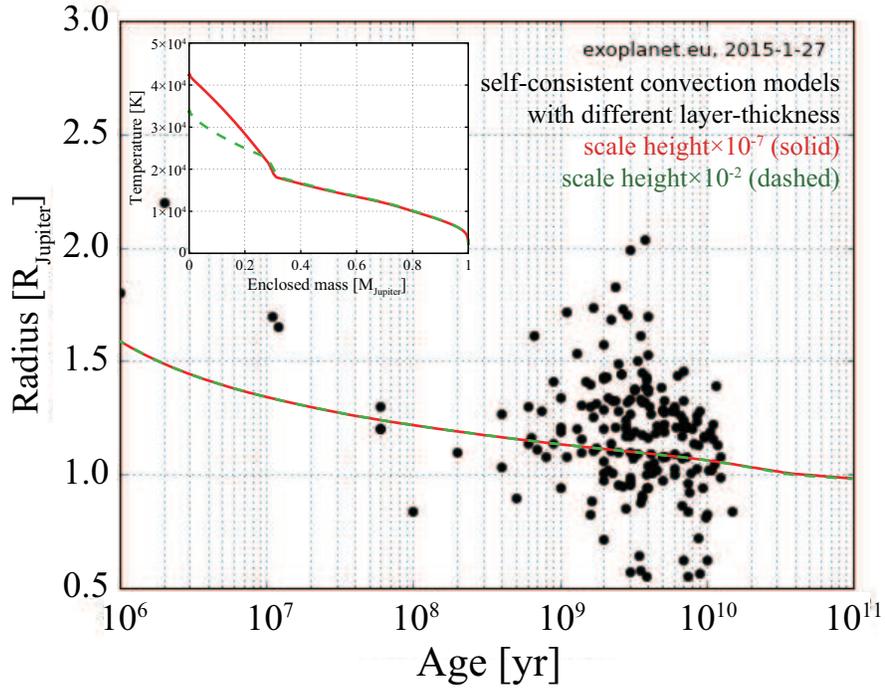}
\caption{
Evolution of the radii of hot Jupiters calculated by using the self-consistent convection model with the different layer thickness.
The results for $l/H_P = 10^{-7}$ (red solid line) and $10^{-2}$ (green dashed line) are shown.
The internal temperature profiles at $30\ {\rm Gyr}$ are portrayed in the inner subset.
\label{fig_Age-R_consistent_l}}
\end{center}
\end{figure}

\clearpage

\begin{figure}
\begin{center}
\includegraphics[scale=1.0]{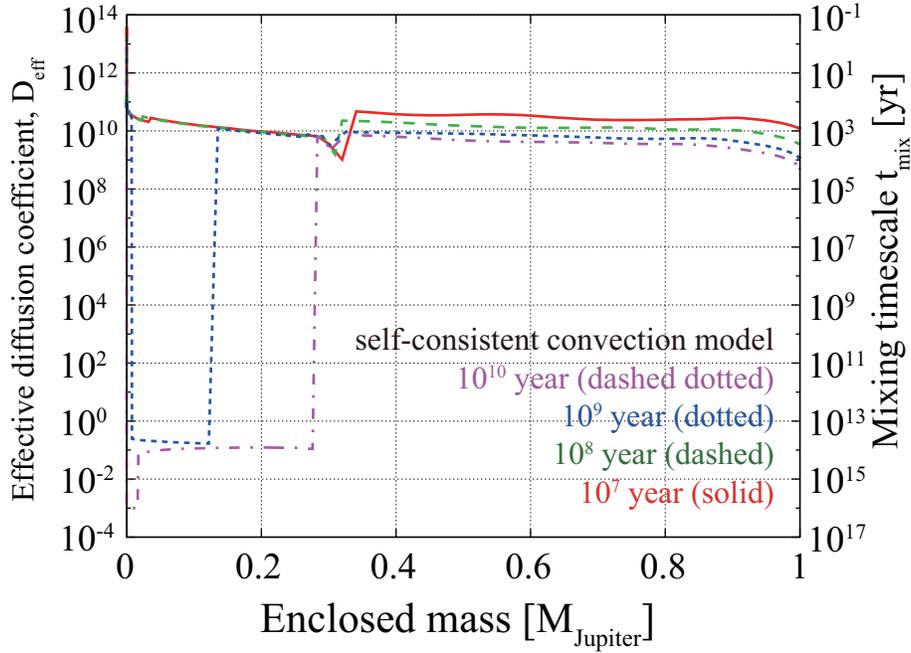}
\caption{
Profiles of the effective diffusion coefficient and the mixing time scale in the interior of the monotonic gradient model calculated with the self-consistent treatment of the convection regimes (the case of red solid line in Fig. \ref{fig_Age-R_consistent}).
The results are shown for $t = 0.01\ {\rm Gyr}$ (red solid line), $0.1\ {\rm Gyr}$ (green dashed line), $1\ {\rm Gyr}$ (blue dotted line), and $10\ {\rm Gyr}$ (magenta dashed-dotted line), respectively.
\label{fig_M-D}}
\end{center}
\end{figure}

\end{document}